\documentclass[letter]{aa} 

\usepackage{graphicx}
\usepackage{dblfloatfix}
\usepackage{lscape}

\usepackage{txfonts}
\begin{document}

   \title{A lensed protocluster candidate at $z=7.66$ identified in JWST observations of the galaxy cluster SMACS0723-7327}
\titlerunning{Protocluster candidate at $z=7.66$}

   \author{N. Laporte
          \inst{1,2}
          \and
          A. Zitrin 
          \inst{3}
          \and
          H. Dole 
          \inst{4}
          \and
          G. Roberts-Borsani 
          \inst{5}
            \and 
          L. J. Furtak
          \inst{3}
            \and 
          C. Witten
          \inst{6}
          }

   \institute{            Kavli Institute for Cosmology,
            University of Cambridge,
            Madingley Road, Cambridge CB3 0HA, UK
            \email{nl408\@cam.ac.uk}
        \and
            Cavendish Laboratory,
            University of Cambridge,
            19 JJ Thomson Avenue, Cambridge CB3 0HE, UK
        \and
            Physics Department, Ben-Gurion University of the Negev, 
            P.O. Box 653, Be’er-Sheva 8410501, Israel
        \and 
            Université Paris-Saclay,   CNRS,  Institut d'Astrophysique Spatiale,
            91405, Orsay, France
        \and 
            Department of Physics and Astronomy, University of California, Los Angeles, 430 Portola Plaza, Los Angeles, CA 90095, USA 
        \and 
        Institute of Astronomy, University of Cambridge, Madingley Road, Cambridge CB3 0HA, UK
             }

   \date{Received 9 August 2022 ; Accepted 3 October 2022}

 
  \abstract
   {According to the current paradigm of galaxy formation, the first galaxies have been likely formed within large dark matter haloes. The fragmentation of these massive haloes led to the formation of galaxy protoclusters, which are usually composed of one to a few bright objects, surrounded by numerous fainter (and less massive) galaxies. These early structures could have played a major role in reionising the neutral hydrogen within the first billion years of the Universe; especially, if their number density is significant. }
   {Taking advantage of the unprecedented sensitivity reached by the \textit{James Webb Space Telescope (JWST)}, galaxy protoclusters can now be identified and studied in increasing numbers beyond $z\geq\ $6. Characterising their contribution to the UV photon budget could supply new insights into the reionisation process.  }
   {We analyse the first JWST dataset behind SMACS0723-7327 to search for protoclusters at $z\geq6$, combining the available spectroscopic and photometric data. We then compare our findings with semi-analytical models and simulations. }
   {In addition to two bright galaxies ($\leq$26.5 AB in F277W), separated by $\sim$11\arcsec and spectroscopically confirmed at $z_{spec}=7.66$, we identify 6 additional galaxies with similar colors in a $\theta\sim20$\arcsec radius around these (corresponding to R$\sim60-90$ kpc in the source plane). Using several methods, we estimate the mass of the dark matter halo of this protocluster,  $\sim$3.3$\times$10$^{11}$M$_{\odot}$ accounting for magnification, consistent with various predictions. The physical properties of all protocluster members are also in excellent agreement with what has been previously found at lower redshifts: star-formation main sequence and protocluster size. This detection adds to just a few protoclusters currently known in the first billion years of the universe. Such $z \ge 7 $ galaxy protoclusters may play an important role in cosmic reionisation.}
   {}


   \maketitle
%
\section{Introduction}

Understanding the formation and evolution of the first population of galaxies a few million years after the Big-Bang is one of the most active topics of current extragalactic astronomy. For decades, instruments have been built to push even further our observational limits. The current most distant and detailed picture of the Universe, the Cosmic Microwave Background (CMB), has been obtained by the \textit{Planck} mission \citep{2020A&A...641A...6P}. It shows that already 380,000 years after the Big-Bang, the matter density in the Universe was inhomogeneously distributed, suggesting that small amplitude density fluctuations were on-going in the early phase of the Universe. These fluctuations grow and eventually the denser regions collapse to form the first bound objects \citep{2011ARA&A..49..373B}. Moreover, the first dark matter haloes undergo a process of fragmentation, suggesting that the most massive galaxies may have formed in overdense regions - so-called protoclusters 
\citep{2010ApJ...719..229G}.  Recent $N$-body simulations and semi-analytic models demonstrate that the first protoclusters may have contributed up to $\sim$50\% of the Cosmic Star Formation Rate Density at $z\sim\ $10 \citep{2017ApJ...844L..23C}. Therefore, determining the number density of $z\geq\ $6 protoclusters could supply new insights into the reionisation process. 

A natural method of identifying 
protoclusters at $z\geq\ $6 is to search for overdensities of photometrically selected dropout galaxies at similar high-redshifts. For example, using \textit{Hubble} Frontier Fields data \citep{2017ApJ...837...97L}, \citet{2014ApJ...795...93Z} identified a likely protocluster of several galaxies at $z\sim\ 8$, one of which was later targeted and found to show UV and FIR emission lines placing it at $z=8.38$ \citep{2017ApJ...837L..21L}. Indeed, searching for galaxies with Lyman-$\alpha$ emission at similar redshifts is another successful route to identify early protoclusters. Since the neutral hydrogen surrounding galaxies at $z\geq\ $6 makes the detectability of this line very low at high-redshift \citep{2017A&A...608A.123D}, a Ly$\alpha$ detection from a galaxy at $z\geq\ $6 suggests that the galaxy sits in a large enough ionized bubble, a result of either high-ionising power of the galaxy itself or the cumulative contribution of several objects at the same redshift \citep{2022arXiv220701629R}. Using the \textit{Hubble} Space Telescope (HST) and deep spectroscopic follow-up campaigns, several groups have identified large ionised bubbles up to $z\sim$8.68 (e.g.  \citealt{2016ApJ...818L...3C,2022A&A...662A.115C}, \citealt{2020ApJ...891L..10T}, \citealt{2021arXiv211207675L},  \citealt{Larson2022ApJ...930..104L}) with the detection of Ly-$\alpha$ in bright galaxies. However, the sensitivity of HST is not sufficient to search for much fainter objects in the local environment of these bright galaxies. 

\begin{figure*}
    \centering
    \includegraphics[width=17cm]{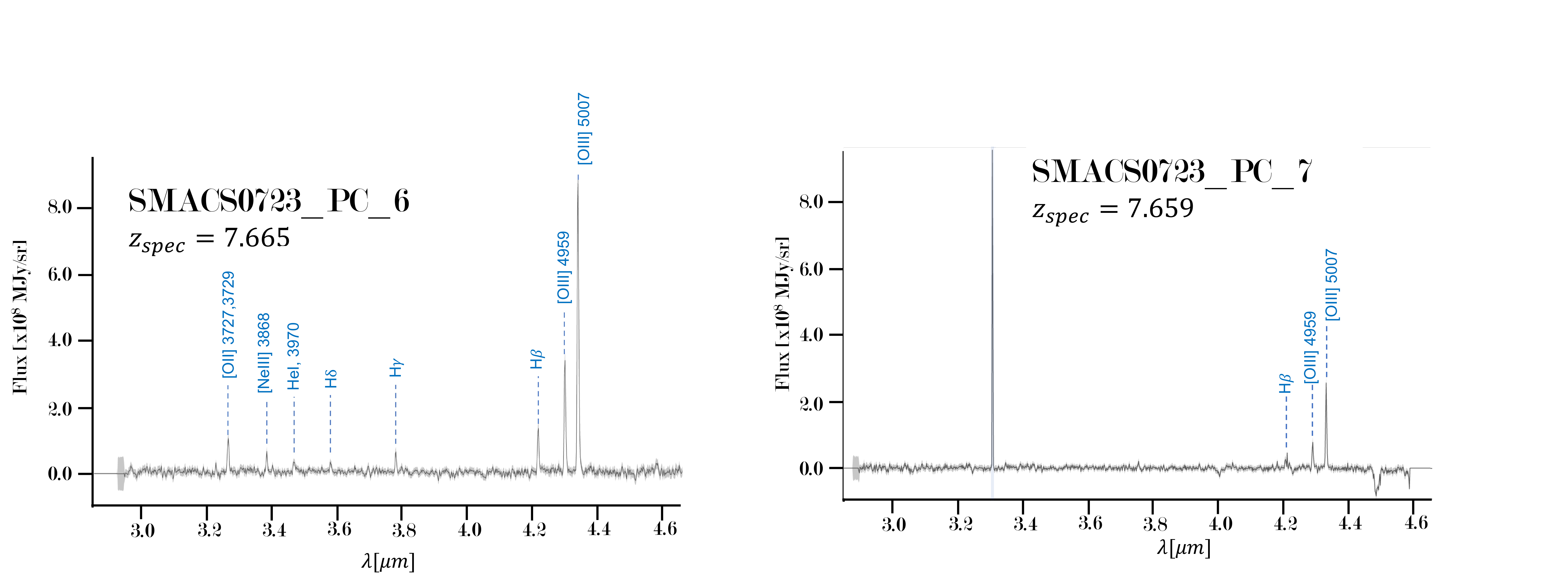}
    \caption{1D NIRSpec spectra of the two galaxies with identical redshift of $z=7.66$. The detected lines are showed with blue dashed lines. A spurious line is seen on the spectrum of SMACS0723\_PC\_6 at 3.31$\mu$m and has been flagged on this figure. Uncertainties on each spectrum are plotted in grey.}
    \label{fig:spectra}
\end{figure*}

The successful launch of the JWST on December 25, 2021 from Europe's Spaceport in French Guyana has opened a new window to study protoclusters. Its unprecedented sensitivity will allow the community to not only identify and spectroscopically confirm bright galaxies, but also fainter galaxies at similar redshifts. On July 12th 2022, the first images and spectra obtained by \textit{Webb} were 
released, and preliminary analysis show that galaxies had  
already formed at $z\geq\ $12 (e.g. \citealt{2022arXiv220712474F}, \citealt{2022arXiv220709434N}, \citealt{2022arXiv220712356D}) reinforcing the idea that a protocluster of galaxies may already be 
in place at $z\geq$10. 

In this paper, we analyse the first dataset released by the JWST behind the lensing cluster SMACS0723-7327 to search for protoclusters at $z\geq$6. In section \ref{sec:search}, we describe our method which leads to the identification of a protocluster at $z=7.66$ using NIRSpec, NIRISS and NIRCam data. Then, we determine the physical properties of protocluster members as well as the global properties of the protocluster (Section \ref{sec:properties}). Finally, in section \ref{sec:implication} we discuss the implication of our findings on the reionisation process.

Throughout this paper we assume a standard $\Lambda$CDM cosmology with parameters from \citet{2020A&A...641A...6P}. All magnitudes are in the AB system \citep{1983ApJ...266..713O}.

\section{Search for protocluster behind SMACS0723}
\label{sec:search}
The first JWST dataset included NIRCam images in F090W, F150W, F200W, F277W, F356W and F444W filters; MIRI images in F770W, F1000W, F1500W and F1800W filters ; NIRSpec spectra in F170LP and F190LP as well as NIRISS spectra obtained with F115W and F200W filters. We first look at the NIRSpec spectra whose integration time is 2.45hrs in each filter. We use the publicly available level 2 data and visually inspect 1D spectra using \texttt{Jdaviz}\footnote{https://doi.org/10.5281/zenodo.6824713}. The spectroscopic redshift is obtained by fitting a list of nebular lines to the brightest detected lines. Two objects among the 35 observed have a similar redshift of $z=7.665$ and $z=7.659$ (herafter SMACS0723\_PC\_6 and SMACS0723\_PC\_7) and show several UV lines such as [OIII]$\lambda \lambda$4959, 5007 and [OII]$\lambda \lambda$3727, 3729 (Figure \ref{fig:spectra}). Our redshift measurements are consistent with what has been previously measured by other groups (e.g. \citealt{2022arXiv220712375C}, \citealt{2022arXiv220708778C}, \citealt{2022arXiv220712338A}, \citealt{2022arXiv220713693K}, \citealt{2022arXiv220714265T}).

Theoretical studies demonstrate that at $z\sim$7 the size of a protocluster is below 10 comoving Mpc, corresponding to $\sim$1 physical Mpc \citep{2017ApJ...844L..23C} . However, the protocluster core, where the large majority of galaxies is expected, is much smaller. For example, \citet{2011Natur.470..233C} identified a protocluster at $z=$5.3, with a protocluster core size of 0.14 physical Mpc. While it appears complicated to entirely cover 
a protocluster at such high-redshift with the first JWST dataset, one can easily search for a protocluster core. In the following, we will search for galaxies with similar colors in a 40\arcsec $\times$40\arcsec box centered on the two galaxies identified at $z=7.66$. Moreover, to increase the number of constraints on the SED, we combine JWST/NIRCam and MIRI data with HST/ACS and WFC3 data.

\begin{figure}
    \centering
    \includegraphics[width=8cm]{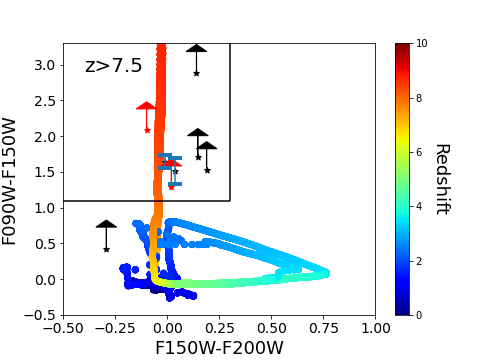}
    \caption{Color-color plot used to select galaxies with color similar to the two galaxies spectroscopically confirmed. Dots show the position of the 8 objects identified in a 40\arcsec $\times$40 \arcsec region. Red dots show the position of the two spectroscopically confirmed galaxies. Lower limits are computed assuming the 2 sigma depth of the F090W image }.
    \label{fig:color}
\end{figure}

\begin{figure}
    \centering
    \includegraphics[width=8cm]{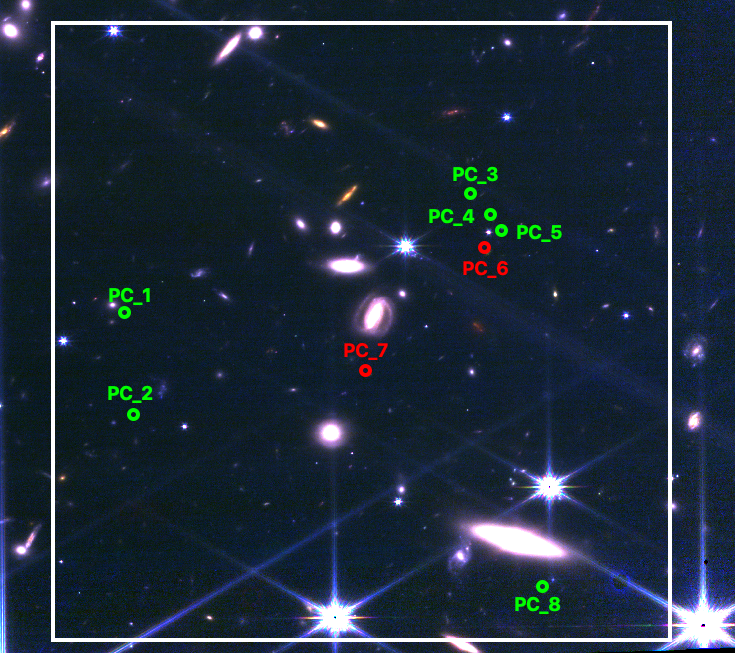}
    \caption{Color image (F090W (Blue) - F150W (Green) - F200W (Red)) of SMACS0723. The white square display the protocluster members search box (40\arcsec $\times$ 40\arcsec), the red 0.3\arcsec radius circles show the position of the two spectroscopically confirmed galaxies at $z=7.66$ and the green circles are at the position of the six other protocluster members candidates identified in this study. }
    \label{fig:FoV}
\end{figure}

To define the color-color criteria needed to select galaxies at $z\sim$7.66, we use BAGPIPES \citep{2018MNRAS.480.4379C} to simulate galaxies whose parameters range from 0.0 to 10 for the redshift, $10^{6}$ to $10^{10} M_{\odot}$ for the stellar mass, -4.0 to -0.5 for the ionisation parameter. The dust attenuation follows \citet{2000ApJ...533..682C} and the dust reddening ranges from $A_v$=0.0 to 2.0mag. IGM attenuation is included and modelled from \citet{2014MNRAS.442.1805I}. We assume 4 different Star Formation History (SFH - constant, burst,  delayed and a combination of a constant with a young burst), and simulate a total of 40,000 galaxies . To account for the depth difference between JWST and HST images, we only defined criteria based on NIRCam filters. We obtain the following criteria to select objects at $z\ge$7 (Figure \ref{fig:color}):  
\begin{align}
    F090W - F150W & > 1.1  \\
    F150W - F200W & < 0.3
\end{align}
We combine these criteria with non-detection criteria ($<2\sigma$) in HST/ACS filters F435W, F606W and F814W and detection criteria ($>5\sigma$) in F150W, F160W and F200W. Colors are measured on psf-matched images degraded to the seeing of F200W (0.07\arcsec), whereas non-detection criteria are measured on original images. We identify 6 more objects in a 40\arcsec $\times$40\arcsec box (Figure ~\ref{fig:FoV}). The photometry of all candidates is reported in Table ~\ref{tab:photometry}.

The SMACS0723 ERO data set also includes NIRISS direct imaging and wide-field slitless spectroscopy (WFSS) in the F115W and F200W filters, which affords sufficient wavelength coverage to observe Ly$\alpha$ at $z>7.2$ or e.g., H$\beta$, OIII]$\lambda$5007 and/or H$\alpha$ for $z\sim2-3$ galaxies. We reduce and analyse the data set as described by \citet{2022arXiv220711387R}, ensuring the modelling and subtraction of neighbouring sources and optimal extraction of 1D spectra using the 2D source profile of the galaxies. One galaxy (SMACS0723\_PC\_8) resides outside of the NIRISS WFSS footprint, while SMACS0723\_PC\_2 and SMACS0723\_PC\_1 are too faint to be detected in the pre-imaging. We visually inspect the resulting spectra of the remaining galaxies (SMACS0723\_PC\_3, SMACS0723\_PC\_4, SMACS0723\_PC\_5, SMACS\_PC\_6 and SMACS0723\_PC\_7) and find no clear emission lines indicative of Ly$\alpha$ or low-$z$ interlopers. Given the spectroscopic verification of SMACS0723\_PC\_6 and SMACS0723\_PC\_7, the lack of emission lines suggest the galaxies have not carved out sufficiently large ionised bubbles for Ly$\alpha$ to escape, while  the absence of strong lines in SMACS0723\_PC\_6 and SMACS0723\_PC\_7 is supportive of such a hypothesis rather than a low-$z$ interloper solution.


\noindent With these numbers in hand 
, we can estimate the overdensity parameter as defined in \citet{2014A&A...568A...1M} :
\begin{equation}
    \delta = \frac{\rho}{\dot{\rho}}-1
\end{equation}
where $\rho$ is the number of objects identified and $\dot{\rho}$ the number of objects expected from the shape of the UV Luminosity Function in our search  box. To obtain the latest value, one first need to estimate the surface explored by our survey by masking all bright objects (stars and obvious low-$z$ galaxies) and by correcting each pixel by the  magnification by the foreground galaxy cluster SMACS0723-7327. The $z\sim$8 UV LF published in \citep{2022arXiv220511526B} is defined between $z\sim$7.5 and $z\sim$8.5, we therefore estimate from the previous effective surface the volume explored within our search box. We find an overdensity parameter of  $\delta=4.0^{+2.4}_{-1.6}$. This value is comparable to what has been computed for protocluster identified at similar redshift, for example $\delta$= 5.11$^{+2.06}_{-1.70}$ for LAGERzOD1 at $z=$6.9 \citep{2021NatAs...5..423H} and $\delta \sim$ 4.5 for the $z\sim$8 overdensity identified in the BoRG survey \citep{2012ApJ...746...55T}. \citet{2022arXiv220802794N} also reported a similar $\sim$4 times overdensity at $z=5.3$ in the CEERS data. 

\section{Physical properties of the protocluster }
\label{sec:properties}

One of the key parameter one has to determine to characterise a protocluster is its total dark matter halo mass. Several methods have recently been used 
in the literature to probe the total halo mass of confirmed protoclusters. In the following, we will apply some of these methods 
to demonstrate that this structure is a convincing protocluster at $z=7.66$.

\begin{figure}
    \centering
    \hspace{-1.0cm}\includegraphics[width=5.cm]{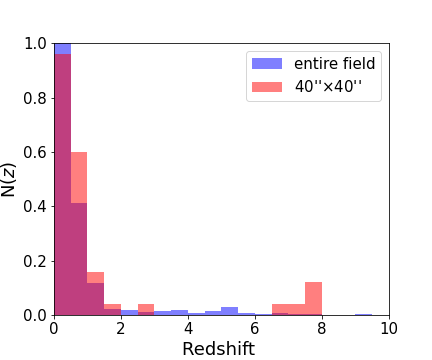}
    \hspace{-0.5cm}\includegraphics[width=5.cm]{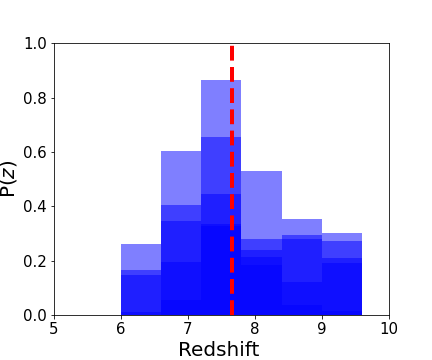}
    \caption{(\textit{Left }) Distribution of the detected objects as a function of redshift in the entire field-of-view (blue) and in the 40\arcsec $\times$40\arcsec box around the two spectroscopically confirmed galaxies at $z=$7.66. The histograms are normalised to 1. A small excess of $z\geq$7 sources  is observed in the protocluster region compared to the entire field-of-view. (\textit{Right }) Redshift probability distribution for the six protocluster members candidates identified in this study. The distribution is compatible with a spectroscopic redshift of $z=$7.66 (vertical dashed red line) .}
    \label{fig:redshift-protocluster}
\end{figure}

Before estimating the total halo mass, one  needs 
to estimate the stellar mass of each candidate protocluster member 
(e.g. \citealt{2020ApJ...898..133L}). We use \texttt{BAGPIPES} \citep{2018MNRAS.480.4379C} and assume several SFH : constant, burst, delayed and a combination of a young burst with a constant SFH. We allow a redshift range of $z\in$[0.0:10.0], a stellar mass ranging from $\log M_{\star} \in$[6.0:12.0] and a reddening range of $A_v \in$ [0.0:2.0].  The best SED-fit is defined as the fit reducing the Bayesian Information Criterion (BIC - see \citealt{2021MNRAS.505.3336L} for more details) and the results of this fitting are presented in Table~\ref{tab:SED-properties}. The photometric redshift of the 6 dropouts identified near the 2 spectroscopically confirmed galaxies is consistent with these galaxies being at $z=$7.66 (Figure \ref{fig:redshift-protocluster} - Right). We apply the same method to obtain the photometric redshift of all objects in the NIRCam field of view. The distribution in redshift of objects in our search box compared to the distribution in the entire field-of-view suggest a small excess of objects at $z\geq$7, consistent with the presence of an overdensity in this region (Figure \ref{fig:redshift-protocluster} - Left). As expected in a protocluster environment  (e.g. \citealt{2015MNRAS.452.2528M},
\citealt{2021MNRAS.501.1803L}, \citealt{2021MNRAS.504.5054A}, \citealt{2022A&A...664A.155G}), 2 members are more massive than the others, namely SMACS0723\_PC\_6 and SMACS0723\_PC\_7, with a stellar mass $\sim$10$^9 M_{\odot}$. 

\begin{figure}
    \centering
    \includegraphics[width=8cm]{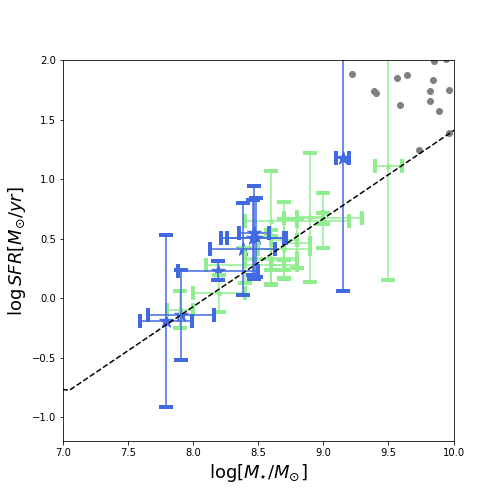}
    \caption{Star-formation main sequence for the 8 galaxies studied in this paper (blue dots) compared with previously published $z\geq$7 sources for which stellar masses are available (green : \citealt{2022arXiv220711135L} ; grey : \citealt{2022arXiv220307392T}). The dashed line shows the parameterisation found by  \citet{2022arXiv220711135L} }
    \label{fig:main-sequence}
\end{figure}

Figure~\ref{fig:main-sequence} shows the position of the 8 galaxies in our sample on a $M_{\star}$ vs SFR diagram compared with previous findings at $z\geq$7 (\citealt{2022arXiv220711135L},\citealt{2022arXiv220307392T}). It confirms that all these sources have properties consistent with what is expected in terms of the star-formation main sequence currently known at $z\geq\ $7. We also expand our search over the entire NIRCam field of view to identify other $z\sim$7.66 objects. 26 objects follow our selection criteria (including the 8 protocluster members candidates discussed in this letter) but no other region in the field shows an overdensity as large as in the protocluster candidate region. Moreover, no candidate has a stellar mass larger than PC\_7 confirming that the two spectroscopically confirmed galaxies are the most massive $z=7.66$ galaxies in this field of view. We also note that the projected distance between the protocluster core discussed in this paper and the closest $z\sim$7.66 galaxy outside of the 40$\arcsec \times$ 40$\arcsec$ box is 0.38 Mpc, suggesting that some of these galaxies may also be related to the protocluster.

    \setcounter{table}{1}
\begin{table*}
\centering
\begin{tabular}{l|cc|cccc|c}
 \hline
 ID  &  RA  &  DEC  &  $z_{phot}$  &  $\log M_{\star}$  &  SFR  &  $A_v$  &  $\mu$ [95\% C.I]\\ 
     &  [deg] & [deg]   &        &  [$M_{\odot}$] & [$M_{\odot}$/yr] &  [mag] &   \\ \hline
 
PC\_1 & 110.823826 & -73.437880 & 6.68$^{+0.04}_{-0.06}$ & 7.79$^{+0.20}_{-0.13}$ & 3.37$^{+0.45}_{-0.18}$ & 0.35$^{+0.30}_{-0.23}$ & 1.90 [1.49 -- 2.06] \\
PC\_2 & 110.820512 & -73.436305 & 7.69$^{+1.88}_{-0.85}$ & 8.19$^{+0.26}_{-0.31}$ & 1.71$^{+0.88}_{-0.77}$ & 0.11$^{+0.11}_{-0.07}$ & 1.79 [1.44 -- 1.92] \\
PC\_3 & 110.845907 & -73.435998 & 7.76$^{+1.19}_{-1.02}$ & 7.96$^{+0.32}_{-0.36}$ & 0.72$^{+0.31}_{-0.25}$ & 0.12$^{+0.15}_{-0.09}$ & 1.73 [1.41 -- 1.83] \\
PC\_4 & 110.846149 & -73.435474 & 7.20$^{+0.55}_{-0.45}$ & 8.57$^{+0.22}_{-0.26}$ & 3.25$^{+1.09}_{-0.83}$ & 0.33$^{+0.14}_{-0.16}$ & 1.70 [1.39 -- 1.80] \\ 
PC\_5 & 110.847412 & -73.435129 & 8.19$^{+0.83}_{-0.57}$ & 8.50$^{+0.20}_{-0.23}$ & 3.23$^{+1.17}_{-1.18}$ & 0.19$^{+0.12}_{-0.11}$ & 1.68 [1.54 -- 1.77] \\
PC\_6$^{\star}$ & 110.844634 & -73.435054 & \textit{7.665}   	 & 8.59$^{+0.19}_{-0.21}$ & 3.55$^{+1.08}_{-1.14}$ & 0.15$^{+0.09}_{-0.07}$ & 1.68 [1.38 -- 1.78] \\
PC\_7$^{\star}$ & 110.834062 & -73.434509 & \textit{7.659}         & 8.95$^{+0.04}_{-0.04}$ & 15.06$^{+2.02}_{-1.89}$ & 1.02$^{+0.03}_{-0.03}$ & 1.68 [1.41 -- 1.79]\\ 
PC\_8 & 110.835151 & -73.429499 & 7.33$^{+1.02}_{-0.09}$ & 8.26$^{+0.16}_{-0.08}$ & 2.61$^{+2.20}_{-0.57}$ & 0.54$^{+0.04}_{-0.06}$ & 1.50 [1.29 -- 1.57]  \\ 
\hline
\end{tabular}
    \caption{ Physical properties computed with \texttt{BAGPIPES} of the 8 protoclusters members candidates identified behind SMACS0723. All the values are not corrected for (the best-fit) magnification. The last column shows the magnification factor estimated from the lens model presented in \citet{2022arXiv220707102P}, assuming a redshift of $z=7.66$ for all objects. \\
    $^{\star}$ : spectroscopically confirmed at $z=7.66$}
    \label{tab:SED-properties}
    
\end{table*}

The individual halo mass can be estimated from the stellar mass of each galaxy using the \citet{2013ApJ...770...57B} relationship. The halo masses of the 8 galaxies range 
from  2$\times$10$^{10}$ to 6$\times$10$^{11}$  M$_{\odot}$ corrected for magnification, with a total protocluster halo mass of 
$M_h$=3.34$^{+0.59}_{-0.50}\times$10$^{11}$M$_{\odot}$ . Another method of estimating 
the halo mass of a 
protocluster is to sum all stellar masses  ($M^{tot}_{\star}$=1.46$^{+0.63}_{-0.41} \times 10^9 $ M$_{\odot}$) corrected for magnification and to convert it into halo mass using the baryonic-to-dark matter fraction measured by \citet{2020A&A...641A...6P}. Following this method, we estimate a total halo mass of 6.52$^{+2.82}_{-1.85} \times 10^{10}$M$_{\odot}$. Finally, we can also determine the halo mass of the most massive galaxy in our sample from its stellar mass, which should dominate the total halo mass of the protocluster. SMACS0723\_PC\_7 has a stellar mass of 5.31$^{+0.54}_{-0.41} \times 10^8$ M$_{\odot}$ corrected for magnification , which converts into a minimum halo mass of 6.86$^{+2.51}_{-2.42}\times$10$^{10}$. 

In an independent analysis, we use the observed stellar-to-halo mass ratios measured by \citet{2022arXiv220310895S} with the COSMOS2020 catalog, which represents the most complete deep galaxy catalog to date \citep{2022ApJS..258...11W}, to compute an additional measurement of the halo mass of our protocluster candidate. We fit the redshift-evolution of the \citet{2022arXiv220310895S} stellar-to-halo mass ratio and extrapolate it out to the redshift of our cluster ($z_{spec}=7.66$) using a Monte-Carlo Markov Chain (MCMC) analysis to rigorously propagate the uncertainties. The resulting halo masses of the 8 cluster member galaxies range from 2$\times$10$^{10}$ to 1$\times$10$^{11}$ M$_{\odot}$, which broadly agrees with the range found using the \citet{2013ApJ...770...57B} relation above. Summing over these halo masses, we find a total halo mass of $M_h$=2.07$\pm1.51\times$10$^{11}$M$_{\odot}$ which is higher than our previous estimates but agrees within the uncertainties ($1\sigma$. Note that with this conversion, the halo mass of the most massive galaxy in the protocluster is also higher than our previous estimate with a halo mass of 10.32$\pm9.78\times$10$^{10}$M$_{\odot}$, though it also has much higher uncertainties. Moreover, \citet{2013ApJ...779..127C} studied the evolution of a Coma-like cluster from $z\sim$7 to $z=0$. Assuming a smooth evolution at $z\geq$7, the expected halo mass of a protocluster at $z=7.66$ is $\sim$3$\times$10$^{11}$M$_{\odot}$. This value agrees well with the total halo mass we have estimated for our protocluster candidate.

Besides the halo mass of the protocluster and the main sequence, another relevant, physical property is its size. As seen in Figure \ref{fig:FoV}, the protocluster constituent galaxies are all found to  
sit within a 40\arcsec $\times$ 40\arcsec box, or within a radius of $\sim20\arcsec$ from the two brighter, central protocluster galaxies. We use the analytic mass model presented in \citet{2022arXiv220707102P} to estimate the magnification of the protocluster. In Table \ref{tab:SED-properties} we list the magnifications of each of the 8 galaxies associated with the protocluster. The magnifications vary between 1.5 and 2, roughly, with a typical $\mu\simeq1.7$ at the center of the protocluster. We do not find a strong shear at the location of the protocluster so that a circle of radius $\sim$20\arcsec encircling the observed protocluster galaxies, is delensed into, approximately, an ellipse of axis ratio 1:1.35, located a few arcseconds closer to the Bright Central Galaxy. Taking the typical magnification into account ($\mu\simeq1.7$), we estimate the size of the protocluster to be about 120 kpc in diameter. This size is consistent with what is expected from \citet{2017ApJ...844L..23C} simulations. It is also similar with what have been observed at $z=5.3$ in \citet{2011Natur.470..233C} and at  $z=6.5$ by \citet{2019ApJ...877...51C}, with protocluster core size of 0.14Mpc and 0.45 Mpc.

\begin{figure}
 
   \hspace{-0.5cm} \includegraphics[width=10cm]{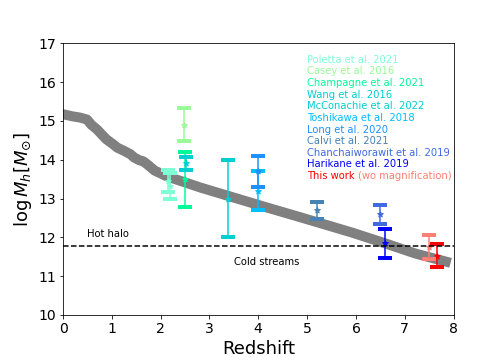}
    \caption{Evolution of protocluster halo mass as a function of redshift. The grey shaded region displays the expected halo mass evolution of a Coma-like cluster (from \citet{2013ApJ...779..127C}) assuming a smooth evolution above $z\geq$7. The dashed line shows the typical threshold mass for a stable shock in a spherical infall, below which the flows are predominantly cold and above which a shock-heated medium is present (from \citealt{2006MNRAS.368....2D}). We overplotted halo mass of recently studied protoclusters at $z\geq$2 from : \citet{2021A&A...654A.121P},  \citet{2016ApJ...824...36C}, \citet{2021ApJ...913..110C}, \citet{2016ApJ...828...56W}, \citet{2022ApJ...926...37M}, \citet{2018PASJ...70S..12T}, \citet{2020ApJ...898..133L}, \citet{2021MNRAS.502.4558C}, \citet{2019ApJ...877...51C} and \citet{2019ApJ...883..142H}. Our halo mass estimate (red dot - the sum of halo masses of individual members) is fully consistent with what is expected for Coma-like cluster. The point in salmon color show the total halo mass without accounting for magnification. It has been sloghtly shifted at lower redshift for clarity purposes. This figure also illustrates the importance to study galaxy protoclusters from Cosmic Dawn to Cosmic Noon.
    }
    \label{fig:evol-proto-cluster}
\end{figure}


\section{Implication for Cosmic Reionisation }
\label{sec:implication}

The transmission and escape of Lyman-$\alpha$ and Lyman continuum photons through the intergalactic medium around a galaxy depends exponentially on the optical depth (\citealt{2004ApJ...613....1F}, \citealt{2000ApJ...530....1M}). As such, even small amounts of neutral Hydrogen will absorb potentially ionizing photons. As a result, smaller and fainter galaxies can only very locally ionize their surroundings, if at all.  This leads to a strong bias towards detecting Lyman-$\alpha$ from brighter and more massive galaxies in the reionization era (e.g. \citealt{2011ApJ...728L...2S}), since typically, only these radiate strongly enough to ionize the surrounding hydrogen and create a sufficiently large bubble to allow these photons to escape (e.g. \citealt{2021arXiv211207675L}, \citealt{2022ApJ...930..104L}).  Hence, smaller galaxies that sit close to these more massive and brighter galaxies, effectively sit in an ionized bubble and their UV photons can travel large distances, to help reionize the Universe (photons that would otherwise 
be absorbed 
in the local vicinity of the galaxies). Therefore, protoclusters may play a significant and important role in the reionization process.

\citet{2017ApJ...844L..23C} demonstrated, using $N-$body simulations and semi-analytical models that protoclusters could have contributed up to $\sim$50\% of the Cosmic Star Formation Rate density at $z=10$. They also show that a protocluster core at $z=$7.66 can itself represent $\sim 10\%$ of the total ionising budget. Furthermore, \citet{2021MNRAS.504.4062M} analyse in their simulations a protocluster with a halo mass similar to the protocluster we report in this letter ($M_h \sim 10^{11} M_{\odot}$), and conclude that the escape fraction for this type of protocluster could reach $f_{esc}\sim$20\%, a \textit{golden number} to explain the end of the reionisation process by $z\sim$6. 

Moreover, the latest constraints on the shape of the UV Luminosity Function at $z\geq$6 show that the number density of galaxies within the first billion years of the Universe is $\leq$10$^{-6}$Mpc$^{-3}$ (e.g. \citealt{2022arXiv220511526B}, \citealt{2015ApJ...810...71F}) which is similar to the number density of  rich clusters at $z=0$ ($\ge$10$^{15}$ M$_{\odot}$ - e.g. \citealt{2010MNRAS.407..533W}). This could suggest that the brightest $z\geq$6 galaxies detected in deep HST surveys may lie in overdense regions with the vast majority of sources well below the detection limit of \textit{Hubble}. If this is the case, and assuming that each $z\geq\ $6  protocluster has an escape fraction of $\sim$20\%, as suggested by simulations, it could open a new route to solve the UV photon deficit observed during the reionisation process. 

Figure~\ref{fig:evol-proto-cluster} demonstrates that SMACS0723\_PC could be seen as the progenitor of a Coma-like cluster ($M_h \ge\ $10$^{15}$M$_{\odot}$), and illustrates the importance of studying galaxy protoclusters from Cosmic Dawn to Cosmic Noon, e.g. in this $M_h-z$ plane. Previous telescopes were only able to detect the brightest members of such structures. Even if recent HST studies start to see overdensities 
near 
extremely bright objects at $z\geq$7 (e.g  \citealt{2021arXiv211207675L}, \citealt{2022A&A...662A.115C}), the density of protoclusters within the first billion years of the Universe is totally unknown. The identification of a protocluster candidate at $z=7.66$ in less than 15hrs with JWST is encouraging. Wider and deeper data are now needed to determine how many bright $z\geq$7 sources lie in overdense regions, which could give new insights into the cosmic reionisation. 

\section{Summary}

In this letter, we report the detection of a protocluster candidate at $z=7.66$ behind the lensing cluster SMACS0723-7327 observed with the \textit{James Webb} Space Telescope. In addition to the spectroscopic confirmation of two $z=7.66$ sources, we identified 6 more galaxies with similar colors whose SED-fitting suggest they may lie at the same redshift, and have physical properties comparable to galaxies at the same redshift. Assuming they are all part of the same structure, we estimate an overdensity parameter $\delta$=4.0$^{+2.4}_{-1.6}$, consistent with previous values found for other protoclusters at $z\geq\ $6. Based on several method, we estimate the total dark matter halo mass of this protocluster candidate to be $M_h=$3.6$^{+13.3}_{2.8} \times$10$^{11}$M$_{\odot}$. This value agrees perfectly with what is expected for progenitors of a Coma-like cluster. Furthermore, the star-formation main sequence at $z \geq 7$ and the estimated size 
are in line with expectations.  

Simulations predict that protoclusters may have played an important role in the cosmic reionisation, with up to a 50\% contribution to the Cosmic Star Formation Rate density. They also suggest that protocluster with the dark matter halo mass of SMACS0723\_PC could have an escape fraction as high as 20\%. The main unknown parameter is their number density. However, the number density of bright galaxies at $z\geq$6 is of the same order of magnitude as the number density of rich clusters at $z\sim$0, suggesting that they may be linked. If this is confirmed with the detection of protocluster candidates around many bright galaxies at $z\geq$6 with the JWST , and given the efficiency at which these early structures ionise the neutral hydrogen, it could give new insight into the reionisation process. 

\begin{acknowledgements}
      We thank Francesco D'Eugenio for providing NIRSpec spectra and Richard Ellis for interesting comments on our manuscript. NL acknowledges support from the Kavli foundation. AZ and LF acknowledge support by Grant No. 2020750 from the United States-Israel Binational Science Foundation (BSF) and Grant No. 2109066 from the United States National Science Foundation (NSF), and by the Ministry of Science \& Technology, Israel. CW acknowledges support from the Science and Technology Facilities Council (STFC) for a Ph.D. studentship. NL and HD acknowledge the \textit{Astr'Auvergne} astronomy festival (and \textit{InfiniSciences} and \textit{Astro-Jeunes} volunteers) as it all started there. This work is based in part on observations made with the NASA/ESA/CSA James Webb Space Telescope. The data were obtained from the Mikulski Archive for Space Telescopes at the Space Telescope Science Institute, which is operated by the Association of Universities for Research in Astronomy, Inc., under NASA contract NAS 5-03127 for JWST. The authors acknowledge the ERO team (program \#2736) for developing their observing program with a zero-exclusive-access period.
\end{acknowledgements}

%
   \bibliographystyle{aa} 
   \bibliography{aanda} 
%
\begin{landscape}
    \setcounter{table}{0}
\begin{table}
\centering
\scriptsize
\begin{tabular}{l|ccccccccccccc}
 \hline
 ID  &  F435W & F606W & F814W & F090W & F105W & F125W & F140W & F150W & F160W & F200W & F277W & F356W & F444W  \\ 
  & ACS/HST & ACS/HST & ACS/HST & NIRCam/JWST & WFC3/HST & WFC3/HST & WFC3/HST & NIRCam/JWST & WFC3/HST & NIRCam/JWST & NIRCam/JWST & NIRCam/JWST & NIRCam/JWST  \\ \hline 
PC\_1	&	>	27.22	&	>	27.84	&	>	26.80	&	29.22	$\pm$	0.47	&	>	27.63			&	>	27.49			&>	27.77			&	27.71	$\pm$	0.13	&>	27.40			&	27.67	$\pm$	0.11	&	27.80	$\pm$	0.24	&	28.19	$\pm$	0.33	&	28.37	$\pm$	0.56	\\
PC\_2	&	>	27.72	&	>	28.47	&	>	27.55	&>	29.93			&	>	27.51			&	>	28.06			&>	27.10			&	28.47	$\pm$	0.15	&>	28.00			&	28.76	$\pm$	0.18	&	29.19	$\pm$	0.48	&	28.71	$\pm$	0.31	&	29.33	$\pm$	0.72	\\
PC\_3	&	>	27.67	&	>	28.21	&	>	27.68	&>	29.69			&	>	27.38			&		27.78	$\pm$	0.54	&	27.00	$\pm$	0.25	&	27.38	$\pm$	0.07	&	26.85	$\pm$	0.25	&	27.19	$\pm$	0.05	&	27.25	$\pm$	0.14	&	27.16	$\pm$	0.12	&	27.48	$\pm$	0.23	\\
PC\_4	&	>	27.59	&	>	27.99	&	>	27.47	&>	29.65			&		27.56	$\pm$	0.51	&	>	26.77			&	26.94	$\pm$	0.28	&	27.19	$\pm$	0.06	&	27.08	$\pm$	0.31	&	27.05	$\pm$	0.05	&	27.30	$\pm$	0.18	&	27.28	$\pm$	0.14	&	27.28	$\pm$	0.18	\\
PC\_5	&	>	27.76	&	>	28.31	&	>	27.63	&	28.21	$\pm$	0.13	&		27.34	$\pm$	0.37	&	>	25.89			&	26.70	$\pm$	0.20	&	26.57	$\pm$	0.03	&	26.53	$\pm$	0.18	&	26.58	$\pm$	0.03	&	26.68	$\pm$	0.08	&	26.75	$\pm$	0.07	&	26.79	$\pm$	0.11	\\
PC\_6	&	>	27.66	&	>	28.23	&	>	27.72	&>	29.73			&		27.46	$\pm$	0.43	&		26.20	$\pm$	0.43	&	26.26	$\pm$	0.14	&	26.02	$\pm$	0.02	&	26.04	$\pm$	0.12	&	25.88	$\pm$	0.02	&	25.74	$\pm$	0.04	&	25.49	$\pm$	0.03	&	24.60	$\pm$	0.02	\\
PC\_7	&	>	27.76	&	>	28.07	&	>	27.69	&>	29.83			&		26.95	$\pm$	0.33	&		27.06	$\pm$	0.24	&	26.54	$\pm$	0.35	&	26.82	$\pm$	0.04	&	26.78	$\pm$	0.20	&	26.92	$\pm$	0.04	&	26.87	$\pm$	0.07	&	26.85	$\pm$	0.07	&	26.09	$\pm$	0.05	\\
PC\_8	&	>	27.70	&	>	28.22	&	>	27.65	&>	28.23			&		OOF			&		OOF			&	OOF			&	27.62	$\pm$	0.19	&	OOF			&	27.60	$\pm$	0.14	&	27.23	$\pm$	1.10	&	27.37	$\pm$	0.33	&	>27.19	\\ 
\end{tabular}
\caption{ Photometry of the selected protocluster members candidates. 2$\sigma$ non-detection are measured at the position of the candidate in a 0.3\arcsec radius aperture. Object PC\_8 is not covered by current HST data, and is marked as Out of the Field (OOF).}
    \label{tab:photometry}
    
\end{table}  
\end{landscape}

\end{document}